\documentclass[longauth]{aa}
\usepackage{txfonts}
\usepackage{graphicx}
\usepackage{longtable}
\usepackage{color}
\usepackage{dcolumn} 
\usepackage{natbib}
\bibpunct{(}{)}{;}{a}{}{,} 
\newcommand{\esnospace}{1ES\,0414$+$009}
\newcommand{\es}{\esnospace\ }
\newcommand{\ergcmsnospace}{\ensuremath{\mathrm{erg}\,\mathrm{cm}^{-2}\,\mathrm{s}^{-1}}}
\newcommand{\ergcms}{\ergcmsnospace\ }
\newcommand{\hessnospace}{H.E.S.S.}
\newcommand{\hess}{\hessnospace\ }

\begin{document}

\title{Discovery of hard-spectrum $\gamma$-ray emission from the BL~Lac object \es}
\small{
\author{HESS Collaboration
\and A.~Abramowski \inst{1}
\and F.~Acero \inst{2}
\and F.~Aharonian \inst{3,4,5}
\and A.G.~Akhperjanian \inst{6,5}
\and G.~Anton \inst{7}
\and A.~Balzer \inst{7}
\and A.~Barnacka \inst{8,9}
\and U.~Barres~de~Almeida \inst{10}\thanks{supported by CAPES Foundation, Ministry of Education of Brazil}
\and Y.~Becherini \inst{11,12}
\and J.~Becker \inst{13}
% ...
}
}
\institute{
Universit\"at Hamburg, Institut f\"ur Experimentalphysik, Luruper Chaussee 149, D 22761 Hamburg, Germany \and
Laboratoire Univers et Particules de Montpellier, Universit\'e Montpellier 2, CNRS/IN2P3,  CC 72, Place Eug\`ene Bataillon, F-34095 Montpellier Cedex 5, France \and
Max-Planck-Institut f\"ur Kernphysik, P.O. Box 103980, D 69029 Heidelberg, Germany \and
Dublin Institute for Advanced Studies, 31 Fitzwilliam Place, Dublin 2, Ireland \and
National Academy of Sciences of the Republic of Armenia, Yerevan  \and
Yerevan Physics Institute, 2 Alikhanian Brothers St., 375036 Yerevan, Armenia \and
Universit\"at Erlangen-N\"urnberg, Physikalisches Institut, Erwin-Rommel-Str. 1, D 91058 Erlangen, Germany \and
Nicolaus Copernicus Astronomical Center, ul. Bartycka 18, 00-716 Warsaw, Poland \and
CEA Saclay, DSM/IRFU, F-91191 Gif-Sur-Yvette Cedex, France \and
University of Durham, Department of Physics, South Road, Durham DH1 3LE, U.K. \and
Astroparticule et Cosmologie (APC), CNRS, Universit\'{e} Paris 7 Denis Diderot, 10, rue Alice Domon et L\'{e}onie Duquet, F-75205 Paris Cedex 13, France \thanks{(UMR 7164: CNRS, Universit\'e Paris VII, CEA, Observatoire de Paris)} \and
Laboratoire Leprince-Ringuet, Ecole Polytechnique, CNRS/IN2P3, F-91128 Palaiseau, France \and
Institut f\"ur Theoretische Physik, Lehrstuhl IV: Weltraum und Astrophysik, Ruhr-Universit\"at Bochum, D 44780 Bochum, Germany
% ...
}

\abstract
{\es ($z$ = 0.287) is a distant high-frequency-peaked BL Lac object, and has long been considered   
a likely emitter of very-high-energy (VHE, $E$ $>$ 100 GeV) $\gamma$-rays
due to its high X-ray and radio flux.
}
{Observations in the VHE $\gamma$-ray band and across the electromagnetic spectrum
can provide insights into the origin of highly energetic particles present in
the source and the radiation processes at work. Because of the distance of the source, the $\gamma$-ray 
spectrum might provide further limits on the level of the Extragalactic Background Light (EBL).}
{We report observations made between October 2005 and December 2009 with \hessnospace, an array of
four imaging atmospheric Cherenkov telescopes. Observations at high energies 
(HE, 100 MeV -- 100 GeV) with the \textit{Fermi}-LAT instrument
in the first 20 months of its operation are also reported. To complete the multi-wavelength
picture, archival UV and X-ray observations with the \textit{Swift} satellite and 
optical observations with the ATOM telescope are also used.}
{Based on the observations with \hessnospace, \es is detected for the first time in the VHE band. 
An excess of 224 events 
is measured, corresponding to a significance of 7.8$\sigma$. The photon spectrum of the source 
is well described by a power law, with photon index of 
$\Gamma_{\rm VHE}$ = $3.45\pm0.25_{\rm stat}\pm0.20_{\rm syst}$. 
The integral flux above 200 GeV is ($1.88\pm0.20_{\rm stat}\pm0.38_{\rm syst}$)~$\times10^{-12}$~cm$^{-2}$~s$^{-1}$.
Observations with the \textit{Fermi}-LAT in the first 20 months 
of operation show a flux between 200 MeV and 100 GeV of (2.3~$\pm$~0.2$_{\rm stat}$)~$\times$~10$^{-9}$~\ergcmsnospace, 
and a spectrum well described by a power-law function with a photon index $\Gamma_{\rm HE}$~=~1.85~$\pm$~0.18.
\textit{Swift}/XRT observations show an X-ray flux between 2 and 10 keV 
of ($0.8-1$) $\times$ 10$^{-11}$~\ergcmsnospace, and a steep spectrum $\Gamma_{\rm X}=(2.2-2.3)$.
Combining X-ray with optical-UV data, a fit with a log-parabolic function 
locates the synchrotron peak around 0.1 keV.}
{Although the GeV-TeV observations do not provide better constraints
on the EBL than previously obtained, they confirm a low density of the EBL, close to the lower limits 
from galaxy counts. The absorption-corrected HE and VHE $\gamma$-ray spectra are both hard and 
have similar spectral indices
($\approx1.86$), indicating 
no significant change of slope between the HE and VHE $\gamma$-ray bands,
and locating the $\gamma$-ray peak in the SED above 1-2 TeV. 
As for other TeV BL Lac objects with the $\gamma$-ray peak at such high energies and a large separation
between the two SED humps, this average broad-band SED represents a challenge for simple one-zone synchrotron 
self-Compton models, requiring a high Doppler factor and very low B-field.
}

\offprints{francesca.volpe@mpi-hd.mpg.de, luigi.costamante@stanford.edu}

\maketitle

\section{Introduction} \label{intro}

The BL Lac object \es was first detected with the \textit{HEAO~1} satellite \citep{gursky78} in the energy range 0.2~keV--10 MeV, and
identified in Einstein Observatory X-ray images \citep{giacconi}. 
On the basis of the original X-ray, optical and radio observations, it was identified as a probable 
BL Lac object by \citet{ulmer83}. 
Located at a redshift of $z$~=~0.287 \citep{halpern91}, \es harbours a super-massive 
black hole of mass $\sim 2 \times 10^9$~$M_\odot$\citep{falomo2003}.    
The host galaxy is classified as elliptical, with 
absolute magnitude $M_{\rm R}=-23.5$ \citep{falomo2003}.
Polarization measurements by \citet{impey_tapia88} confirmed the classifications of this source as a BL Lac object.  
According to the classification scheme of \citet{padovani_giommi95}, \es belongs to the class of high-frequency-peaked BL Lac (HBLs), 
objects with a synchrotron-emission peak located at UV/soft-X-ray
frequencies, or equivalently, sources for which the X-ray emission is dominated by synchrotron radiation.

There are several archival measurements of the spectral properties of \es in X-rays taken using 
\textit{ASCA} \citep{ASCA}, \textit{ROSAT} \citep{ROSAT}, and \textit{BeppoSAX} \citep{BEPPOSAX}.
Observations with \textit{EXOSAT} \citep{giommi90,sambruna94} showed evidence of temporal variability and
strong spectral variations, at times having an X-ray photon index of $\Gamma_{\rm X}<2$.
During these states \es is characterized by a SED 
typical of ``extreme" BL Lac objects 
\citep[i.e. objects with the synchrotron peak above a few keV,][]{costa_extreme}.
In the VHE $\gamma$-ray domain, the data from the HEGRA experiment were used to derive an upper limit on the flux 
for this source corresponding to 13.5 $\times 10^{-12}$~cm$^{-2}$~s$^{-1}$ above 910 GeV \citep{HEGRA_UL}. 
\es was considered a good candidate for VHE emission by \citet{costa_ghisellini},
on the basis of a high X-ray and radio flux, and its detection was yet more
likely after blazar $\gamma$-ray spectra indicated 
a low intensity of the diffuse EBL \citep{nature}.

Based on these VHE estimates, observations of \es with \hess began in 2005 and continued until 2009,
and significant observing time was dedicated to  
its detection, because its high redshift made this source a potentially interesting candidate for EBL studies.
In Sect.~\ref{HESS_OBS} the analysis of the 5-year data set on \es collected 
by the \hess collaboration is reported. 
In the high-energy $\gamma$-ray domain, emission from \es was first detected by
the Large Area Telescope (LAT) on the \textit{Fermi} $\gamma$-ray space telescope \citep{REF::LAT_INSTRUMENT}. 
\es was detected with a significance of $TS = 2\Delta\log\mathcal{L} > 25$
 in the first 11 months of the \textit{Fermi}-LAT operation and
is listed in the first year \citep{REF::ONEFGL} and 
second year \citep{REF::2FGL} \textit{Fermi} catalogues.
In Sect.~\ref{MWLAna} the 20 month \textit{Fermi}-LAT observation of \es 
is reported. The $\gamma$-ray observations 
have been complemented with measurements in X-ray and UV, carried out
using the \textit{Swift}/XRT and UVOT instruments 
\citep{xrt,Roming}. These, together with the optical measurements from the ATOM telescope \citep{Hauser04}
are presented. 
Constraints on the EBL are discussed in Sect.~\ref{spectral_fits}, and
finally, the spectral energy distribution (SED) of \es is presented in Sect.~\ref{Discussion}.

In this paper the luminosity distance of the source, located at a redshift 
$z$ = 0.287 
is computed with the standard $\Lambda$CDM cosmological model ($H_0$~=~71~km$/$s$/$Mpc, $\Omega_{\Lambda}$~=~0.73, $\Omega_{M}$~=~0.27), yielding a luminosity distance $D_L$~=~1469.3~Mpc. 

\section{VHE $\gamma$-ray observations with \hess }\label{HESS_OBS}

\hess is an array of four imaging atmospheric Cherenkov telescopes situated
in the Khomas highland of Namibia ($23^\circ16\arcmin18\arcsec$ south,
$16^\circ30\arcmin00\arcsec$ east), at an elevation of 1800 metres above sea
level \citep[see][]{HESS_Crab}. The \hess array is sensitive to $\gamma$-rays above
$\sim$ 100~GeV, and commonly achieves an event-by-event angular resolution 
of $\lesssim$0.1$^{\circ}$ and a relative energy resolution of $\sim$15\%.

The BL Lac \es was observed every year between 2005 and 2009. After run-quality
selection the data set comprises 73.7 hours (live time) of observations,
of which 67.5 hours were taken with four telescopes. Observations have been carried out at 
zenith angles of 22$^\circ$ to 41$^\circ$ with a mean value of 26$^\circ$. 
The pointing offset from \es was 0.5$^\circ$. 
The data were analysed with the Model Analysis \citep{denaurois}, 
in which shower images of all triggered telescopes are compared 
to a precalculated model by means of a log-likelihood minimisation. 
The analysis has been cross-checked with a multivariate analysis 
\citep{ohm} using an independent calibration, which yielded consistent results.

Two sets of cuts were used. For the source-detection analysis {\em standard cuts}
were used, including a minimum image charge of 60 photoelectrons 
(energy threshold E$_{\rm th}$~=~220~GeV,
defined here as the energy at 
which the effective area falls to 10\% of its maximum value).
{\em Loose cuts}, with a smaller charge cut of
40 photoelectrons, result in a decreased signal-to-background
ratio but have the advantage of higher photon statistics and a lower threshold of
150 GeV.
Since \es is a very faint and distant source, {\em loose cuts} were used for 
the spectral analysis and light curve generation \footnote{In this case the {\em loose cuts} 
almost double the number of excess events resulting in a poorer background rejection.}. Therefore, 
in this text all the fluxes and light curves shown are derived with {\em loose cuts}. 

\begin{figure}[!t]
  \centering \includegraphics[width=9cm]{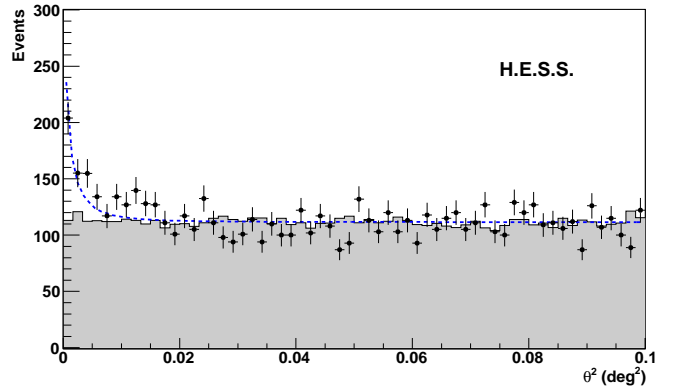}
  \caption{The distribution of the squared angular distance ($\theta^2$) for the ON-source events (crosses) 
   and the OFF-source events (shaded). 
The dashed blue line represents a PSF profile for a point-like source estimated by Monte Carlo simulations.}
  \label{Theta2_HESS}
\end{figure}

A significant excess of 224 events (7.8$\sigma$) from the direction of \es was found in the total data
set using {\em standard cuts}. A year-by-year summary of the observations and of the corresponding results is shown in Table~\ref{YearWiseHESS}. 
The background subtracted distribution of  
the square of the angular difference between the reconstructed shower position and 
the nominal source position ($\theta^2$)
is shown in Fig.~\ref{Theta2_HESS}. 
The {\em Reflected-Region} method \citep{berge07} was used for the 
definition of the ON-source and OFF-source data regions.
The $\theta^2$ distribution is consistent with that of a point-like source, 
as shown in Fig.~\ref{Theta2_HESS}

\begin{table*}[t]
\begin{center}
\caption{Summary of \hess observations of \es for each year. 
The MJD of the first and last night of the observation 
in each epoch between 2005 and 2009, the live-time, the number of ON- and OFF- source events, 
the excess and its significance are reported. 
These events are obtained with {\em standard cuts} and 
for all the observation the normalization factor between the OFF and ON exposures is 12.6.
The last three columns list the integral flux above 200 GeV, 
the $\chi^2$, number of degrees of freedom (ndf) and the $\chi^2$ probability $P$($\chi^2$) for a constant fit to the light curves binned by night within each year (first five rows), or yearly (last row) within the total observing period. 
\label{YearWiseHESS}}
\begin{tabular}{*{11}{c}}
   \hline\hline
Epoch   & MJD$_{\rm start}$   &  MJD$_{\rm stop}$  &  Live-time[h] &  ON &  OFF & Excess & Sig[$\sigma]$ & $F$($E>$200 GeV) $^{(*)}$ & $\chi^{2}$$/$ndf $^{(**)}$ & $P$($\chi^2$) $^{(**)}$\\ 
&  & & & & & & & 10$^{-12}$~cm$^{-2}$~s$^{-1}$& & \\
\hline
2005   & 53646.06 & 53706.97 & 15.2 & 162 & 1474 & 44.9 & 3.8 & 2.16$\pm$0.41 & 20.5$/$16& 0.20 \\
2006   & 54024.03 & 54120.82 & 11.4 & 189 & 1834 & 43.3 & 3.3 & 1.19$\pm$0.51 & 10.5$/$12&0.57 \\
2007   & 54406.98 & 54475.84 & 12.9 & 150 & 1541 & 28.1 & 2.4 & 1.73$\pm$0.46 & 13.9$/$11&0.24 \\
2008   & 54736.09 & 54750.10 & 9.1 & 117 & 1160  & 25.1 & 2.4 & 1.37$\pm$0.62 & 2.5$/$8&0.96 \\ 
2009   & 55067.14 & 55158.97 & 25.2 & 289 & 2602 & 82.8 & 5.2 & 1.86$\pm$0.41 & 31.0$/$28&0.32 \\
   \hline

Total &  & & 73.7 & 907 & 8611& 224.3& 7.8& 1.84$\pm$0.20 & 2.66$/$4&0.62 \\
   \hline
\end{tabular}
\end{center}
 \begin{list}{}{}
 \item[$^{(*)}$] Derived with {\em loose cuts} and with the spectral index fixed to the global value $\Gamma_{\rm VHE} = 3.45$. 
 \item[$^{(**)}$] Night-by-night and year-by-year light curves are derived with {\em loose cuts}. 
  \end{list}
\end{table*}

The fit of the uncorrelated excess map with a two-dimensional point-like function convolved 
with the instrument point spread function PSF  
(68\% containment radius of 0.064$^{\circ}$ for {\em standard cuts}) also indicates that the source is consistent 
with being point-like and is located at ($\alpha_{\rm J2000}$=$4^{\rm h}16^{\rm m}52.96^{\rm s}\pm0.10^{\rm s}_{\rm stat}\pm0.10^{\rm s}$, $\delta_{\rm J2000}$=1$^{\circ}5'20.4^{\rm ''}\pm15^{\rm ''}_{\rm stat}$).
This position is consistent with the nominal position 
($\alpha_{\rm J2000}$=$4^{\rm h}16^{\rm m}52.8^{\rm s}$, $\delta_{\rm J2000}=1^{\circ}5'24^{\rm ''}$) 
reported by \citet{ulmer83}.

\begin{figure}[!]
\centering \includegraphics[width=9cm]{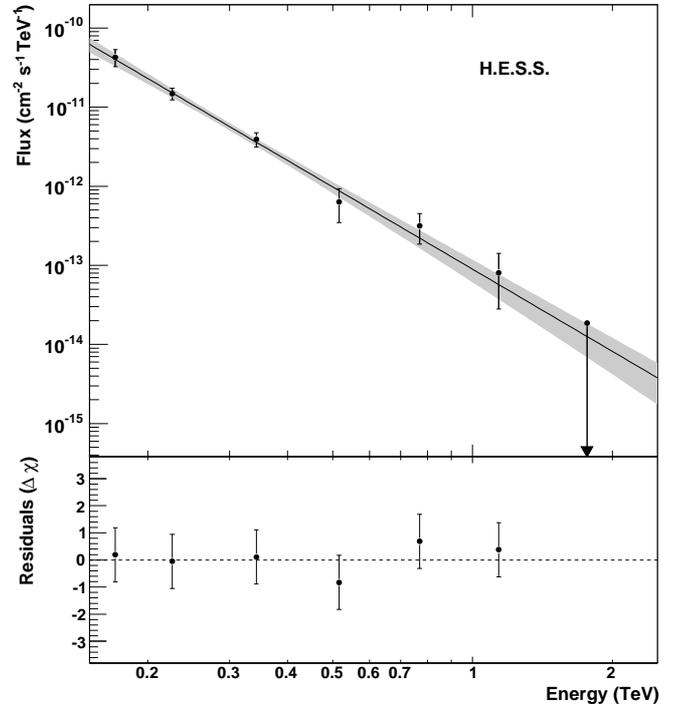}
   \caption{Differential energy spectrum of \esnospace. The shaded band
corresponds to the range of the power-law fit (68$\%$ confidence level), taking into account statistical errors. 
The lower panel shows the residuals of the fit; the difference in each
energy bin between the expected and observed number of excess events, 
normalized to the uncertainty on the latter. 
The 68\% confidence level upper limit is calculated 
with the method of \citet{Feldman}.
}
   \label{DiffSpectrum_HESS}
 \end{figure}

Fig.~\ref{DiffSpectrum_HESS} shows
the differential energy spectrum of the VHE $\gamma$-ray emission 
above the energy threshold of $\sim$ 150 GeV using the $\sim$530 excess events 
obtained with {\em loose cuts}.
The spectrum is obtained using a forward-folding maximum-likelihood method,
described by \citet{ForwardFolding}. 
The spectrum is compatible with a power-law distribution (d$N$/d$E \sim E^{-\Gamma}$)
with a photon index of $\Gamma_{\rm VHE} =3.45 \pm 0.25_{\rm stat} \pm 0.20_{\rm syst}$; 
the differential flux at the decorrelation energy of 305~GeV is
($5.70\pm0.62_{\rm stat}$)~$\times10^{-12}$~cm$^{-2}$~s$^{-1}$~TeV$^{-1}$. 
In Fig.\ref{DiffSpectrum_HESS} the confidence band of the power-law fit for \es is illustrated, 
clearly steep in appearance and affected by EBL absorption.
The spectral points (with 1$\sigma$ statistical errors) 
in the upper panel in Fig.~\ref{DiffSpectrum_HESS} are derived from the residuals in the different 
energy bins shown in the bottom panel in the same figure. 
The integral flux above 200 GeV is $F$~=~($1.88\pm0.20_{\rm stat}\pm0.38_{\rm syst}$) $\times10^{-12}$~cm$^{-2}$~s$^{-1}$ and it corresponds to $\sim0.6\%$ of the Crab Nebula flux above the same energy threshold \citep{HESS_Crab}.

The light curves for $E$~$>$~200~GeV with different time samplings have been derived with {\em loose cuts}, assuming 
the spectral index $\Gamma_{\rm VHE} = 3.45$ obtained by the maximum-likelihood fit 
(see the top panel in Fig.~\ref{LC_MWL}).
No evidence of variability is found on different timescales. 
Accounting for both statistical and systematic (20\%) uncertainties on the flux points, 
the fit of the night-by-night light curve with a constant value yields  $\chi^2=81.1$ (for 79 degrees of freedom)
with a corresponding probability P=0.41 and a 99\% confidence level upper limit
on the fractional variability $F_{\rm var}$~$<$~0.40 with $F_{\rm var}$ defined as in \citet{Vaughan03} 
(calculated with the method described in \citealt{Feldman}).
The fit of the monthly light curve 
yields a $\chi^2=18.73$ for 12 degrees of freedom,
with a corresponding probability of 0.1 and upper limit $F_{\rm var}<$~0.76.
The same conclusion is found on a yearly timescale, with a probability of a constant flux $P$=0.62 and 
99\% confidence level upper limit $F_{\rm var}<$ 0.32.

Table~\ref{YearWiseHESS} shows the measured integral flux for each year the source was observed.
Also reported are the $\chi^2$ values for a fit of a constant when the flux is binned by nights within each year and by years within the total observation. 
As the $\chi^2$ probability for each fit is greater equal 
than 0.2 there is no evidence for variability in this data set. 

\section{Multi-wavelength observations} \label{MWLAna}

\subsection{Gamma-ray observations with the \textit{Fermi} LAT}\label{FERMI_OBS}

The LAT is a pair-conversion telescope, and is the primary instrument
of the \textit{Fermi} satellite launched in June 2008. The
\textit{Fermi}-LAT instrument, described in detail in
\citet{REF::LAT_INSTRUMENT}, detects $\gamma$ rays with energies
between 20\,MeV and $>$300\,GeV. The bulk of the \textit{Fermi}
observational program is dedicated to a survey, in which the full
$\gamma$-ray sky is covered every 3 hours.

The spectrum of \es in the 100\,MeV to 100\,GeV energy
band\footnote{The analysis has been restricted to the energy range
  over which the response of the instrument is best understood.}  was
derived from LAT data taken between August 4, 2008 and May 17, 2010
(651 days). To minimize background contamination, only events with a
high probability of being $\gamma$ rays, known as ``DIFFUSE'' events,
having an angle of less than 105$^\circ$ to the Zenith, were
retained. The spectrum was evaluated using an unbinned maximum
likelihood approach \citep{REF::MATTOX_LIKE} in which the events
originating within a region of interst (ROI) of 10$^\circ$ around \es
were modelled using diffuse Galactic and isotropic components and 17
point sources with power-law spectra. The diffuse components are
tabulated in the standard files \textit{gll\_iem\_v02.fit} and
\textit{isotropic\_iem\_v02.txt}, respectively; the overall
normalization of each component was allowed to vary in the fit. The
point sources include a test source at the location of 1ES~0414+009,
all sources from the first \textit{Fermi} catalogue
\citep{REF::ONEFGL} within 12$^\circ$ of \es and three additional
sources found in a dedicated search for additional $\gamma$-ray
emission in the ROI. These additional sources are weak, with detection
significances between $TS=24$ and $TS=42$ (approximately $4\sigma$ to
5.6$\sigma$), and all lie relatively far from \es, the closest being
separated by $>5.2^\circ$\footnote{These additional sources are not
  associated with any point sources in 2FGL, however one is
  approximately two degrees from a 2FGL source marked as
  \textit{confused}. It seems likely that they are the result of
  deficits in the galactic diffuse model that were eliminated in the
  improved model use in 2FGL.}. For the brighter background sources
the power-law spectral assumption was tested by evaluating a log
parabolic model, which was found not to significantly improve the
fit. In total the model had 30 free parameters\footnote{The spectral
  parameters for point sources formally outside the 10$^\circ$ ROI
  were fixed at their catalogue values \citep{REF::ONEFGL}. They were
  included in the model to account for the effects of the relatively
  large PSF at low energies.}. The maximum likelihood optimization was
performed using the standard \textit{Fermi} ScienceTools package
(v9r19p0), using the P6\_V3\_DIFFUSE instrument response functions
(IRFs).

The null hypothesis, that no source is present at the location of
\esnospace, is rejected with $TS = 67.7$ for 2
degrees of freedom, or approximately $7.9\sigma$. The best-fit power
law has an integral flux of 
$(4.3 \pm 2.2_{\rm stat}$$^{+1.08}_{-0.88\,\rm syst})\times
10^{-9}$~cm$^{-2}$~s$^{-1}$ (100~MeV to 100~GeV) and an index of
$\Gamma_{\rm HE} = 1.85 \pm 0.18$. The differential flux at the
decorrelation energy of 3.6\,GeV is $(5.0 \pm 1.1)\times
10^{-14}$~cm$^{-2}$~s$^{-1}$~MeV$^{-1}$. The systematic errors on the
index, estimated using the bracketing IRF method
\citep{REF::FERMICRAB}, in which the IRFs are modified to give the
greatest change in the index of the best-fit power law, are
$\Gamma_{\rm HE, syst}=^{+0.10}_{-0.12}$. 

The most energetic photon
detected within a region of $0.5^\circ$ around the source was
reconstructed with an energy of $E=19.7$\,GeV. The energy resolution
is less than $10\%$ for photons with $E=20$\,GeV.

Variability in the flux of $\gamma$ rays on monthly timescales was
evaluated by producing a light curve with 32.5 day bins (second panel
of Fig.~\ref{LC_MWL}, see \citet{REF::ONEFGL} for details of
methodology). Comparing the likelihood of a model in which the flux in
each time bin is free, to one where it is assumed constant gives a
difference of $TS_{VAR}=2\Delta\log\mathcal{L}=19.2$ for 19 extra
degrees of freedom (DOF). The probability of getting a value of
$TS_{VAR}>19.2$ by chance is $P=0.44$, assuming the theorem of
\citet{Wilks} is applicable, and we conclude that there is no
evidence for variability on monthly timescales. 
Since there is a
relatively large number of upper limits in the 32.5-day light curve,
we rebinned the data to produce a 130-day
light curve which is also shown in Fig.~\ref{LC_MWL} (blue points). Applying the
same variability test gives $TS_{VAR}=1.44$ for 4 DOF, with a chance
probability of $P=0.84$, again consistent with no variability.

\subsection{X-ray observations with the \textit{Swift}/XRT}\label{XRAY_OBS}
%

Archival \textit{Swift} observations within the timespan of the 
overall \hess observing campaign were analysed.
\textit{Swift} performed three snapshot observations in 2006, on October 21 (1138~$\rm s$), 
22 (226~$\rm s$) and 23 (553~$\rm s$),  and three in 2008, two on January 29 
(1008 and 993~$\rm s$) and one on February 4 (2227~$\rm s$).

The XRT data were processed with standard procedures using the
FTOOLS task XRTPIPELINE (version 0.12.4). Only data taken in photon counting (PC) 
mode were considered, given the low rate of the source ($<0.5$ counts/s in the 0.3-10 keV range)
and because the exposures in Window Timing mode were extremely short.
Source events were extracted in the 0.3$-$10 keV range within a circle with a
radius of 20 pixels (47\arcsec), while background events were extracted from both circular 
and annular regions around the source, free of other sources.

The spectra were fitted with a single power-law model with 
Galactic absorption $N_{\rm H}$ fixed at  $9.15\times10^{20}$~cm$^{-2}$, 
as obtained by dedicated 21~cm observations \citep{elvis99}.
No significant flux or spectral variability was found within the 2006 or Jan 2008 data sets
(the fit with a constant gives $\chi^2$/ndf of 4.9/10  and 10.5/7, respectively,
with a probability of 0.90 and 0.16).
The corresponding exposures were thus summed, 
obtaining a total of three data sets for further analysis:
the first covering the period October 21-23, 2006 (1916~$\rm s$),
the second for January 29 (2001~$\rm s$), and the third for February 4, 2008 (2227~$\rm s$).

To test the possible presence of spectral curvature, broken power-law and log-parabolic
models were compared to the simple power-law hypothesis.
The three spectra are well fitted by a single power law within the available statistics,
with steep indices around  2.2--2.3 (see Table~\ref{swift}).  
Log-parabolic models do not significantly improve the $\chi^2$ (F-test$\sim$85-95\%).
Variations among the different epochs are modest:
both flux and spectra are similar between October 2006 and January 2008,
while on Feb 4 the flux increases by $\sim$15\%,
but consistent within 1 sigma with other two epochs,
with a slightly harder spectrum by $\Delta\Gamma\simeq 0.15\pm0.10$. 
The light curve for the XRT observations is shown in Fig.~\ref{LC_MWL}.

The \textit{Swift} Burst Alert Telescope \citep[BAT,][]{bat} data from 2004 to 2010 
were also analysed,  following the procedure described in \citet{Ajello2008}.
The source presented only a marginal signal of 2.5$\sigma$
in the 14-30 keV band. The corresponding 99\% upper limit on the flux
is reported in the SED in Fig.~\ref{sed}, for reference.

\begin{table*}
\begin{center}
\caption{Spectral parameters and fluxes of the Swift data in the three epochs.}
\label{swift}
\begin{tabular}{lcccc}
   \hline
Quantity  &  21-23 Oct 2006  &  29 Jan 2008  &  4 Feb 2008 &  Units     \\ 
  &       54029.04-54031.04 & 54494.23 & 54500.21 &  MJD\\
   \hline
\multicolumn{5}{c}{Single power-law fit, XRT data}  \\ 
   \hline 
$\Gamma$              & $2.30\pm0.08$ & $2.33\pm0.08$ & $2.18\pm0.07$ &  \\
$F_{\rm 2-10 keV}$    & $8.41\pm0.70$ & $8.92\pm0.70$ & $10.3\pm0.7$  & $10^{-12}$ \ergcms  \\
$\chi^{2}_{r}/{\rm ndf}$        & 1.06$/$42      & 1.00$/$22      &  0.80$/$43   &  \\
 \hline
\multicolumn{5}{c}{UVOT filters, observed magnitudes$^{\rm (a)}$} \\
\hline 
V      & $16.56\pm0.03$ & - & - & \\
B      & $16.97\pm0.03$ & - & - & \\
U      & $16.06\pm0.03$ & - & $16.51\pm0.03$ & \\
UVM2     &   -            & $16.56\pm0.03$ & - & \\
 \hline
 \multicolumn{5}{c}{Log-parabola$^{\rm (b)}$ fit, combined UVOT $+$ XRT data}  \\
 \hline 
UVOT filters          & B, U          &  UVM2           &  U             & \\
$\Gamma_{\rm 1 keV}$      & $2.28\pm0.07$   &  $2.29\pm0.08$  & $2.16\pm0.06$    &  \\
$b$                   & $0.12\pm0.03$   &  $0.17\pm0.03$  & $0.10\pm0.03$    & curvature \\
$E_{\rm peak}^{\rm (c)}$        & $0.071\pm0.007$ &  $0.13\pm0.02$  & $0.17\pm0.06$    & keV \\
$F_{\rm 2-10 keV}$    & $7.83\pm0.70$          &  $8.39\pm0.70$         & $9.96\pm0.70$      & $10^{-12}$ \ergcms \\
$\chi^{2}_{r}/{\rm ndf}$       & 1.01$/$43      &  0.84$/$23     & 0.79$/$44       &  \\
   \hline
\end{tabular}
\end{center}
\scriptsize
 \begin{list}{}{}
  \item Errors are reported at 90\% confidence level for 1 parameter ($\Delta\chi^2=2.71$).
  \item[$^{\mathrm{(a)}}$] Statistical errors from {\tt uvotsource} analysis. 
  \item[$^{\mathrm{(b)}}$] Log-parabola model is defined as: d$N$/d$E \propto E^{-\Gamma+b\,\log(E)}$
  \item[$^{\mathrm{(c)}}$] Value of the peak energy in the SED, obtained by fitting the same data with 
  a different parameterization of the log-parabolic model, for which the free parameters are
  $E_{\rm peak}$ and $b$ \citep[details in][]{tram07}.
 \end{list}
\end{table*}

\subsection{Optical and UV observations with the \textit{Swift}/UVOT}\label{UVOT_OBS}

The UVOT instrument \citep{Roming} took exposures
in different filters during the XRT pointings, namely
V (547\,nm), B (439\,nm), and U (347\,nm) in October 2006, UVM2 (225\,nm) in January 2008 and U in February 2008 
(the observations are summarised in Table~\ref{swift}).
Photometry of the source was performed using the UVOT software 
in the HEAsoft 6.9 package.
Counts were extracted from the standard aperture of 5$"$ radius 
for all single exposures and all filters, while the background
was carefully estimated from different positions more than 27$"$ away from the source. 
Count rates were then converted to fluxes using the standard photometric 
zero points \citep{Poole2008}.

The fluxes were de-reddened for Galactic absorption
using the extinction curve from \citet{Cardelli94} with the updates from \citet{odonnell96},
assuming $R_{\rm V}[=A_{\rm V}/E(B-V)] = 3.1$. This is the average value for the Galactic diffuse 
interstellar medium.
For the line of sight of \esnospace, a value of $A_{\rm B}=0.507$ was used
\citep{schlegel}, with $E(B-V) = 0.117$~mag.

The host galaxy of \es is elliptical, of total R magnitude 17.49 
and half-light radius 4.7$\arcsec \pm0.5\arcsec$  \citep[from HST snapshot observations,][]{scarpa2000}.
To isolate the flux of the active core 
the UVOT fluxes were corrected for the contribution of the host galaxy.
The wavelength-dependent correction was determined using a template
for elliptical galaxies \citep{silva98}, rescaled to the host-galaxy flux in the R band, 
and accounts for the given apertures.

As in the X-ray band, no significant variability was found in the UV band
in October 2006, nor in January 2008  (see fourth panel in Fig.~\ref{LC_MWL},
where fluxes are reported without extinction correction).
The individual frames were thus co-added, and the results are given in Table \ref{swift}.
A decrease by 0.5 mag can be noted between 2006 and February 2008,
in the U filter (the only filter in common).
This behaviour contradicts what observed in the X-ray band, where the corresponding
XRT spectrum showed an increased flux and harder spectrum.
This is indicative of a shift of the SED peak towards higher energies,
which is confirmed by the simultaneous UVOT-XRT spectral fitting (see Sect.~\ref{uvot_xrt_fits}).

\subsection{Optical observations with ATOM}\label{ATOM_OBS}

The 75\,cm optical telescope of the \hess collaboration, ATOM, 
is located next to the Cherenkov telescopes at the \hess site in Namibia. ATOM has
been monitoring \es since Nov 2006 with a sampling of $\approx$2
observations per week during its visibility from July to March.

During the \hess observations of \esnospace, the sampling was
increased to obtain truly simultaneous data. The observations of this source were
mostly carried out in B and R band (440\,nm and 640\,nm). On some of the
nights, observations in V and I bands (550\,nm and 790\,nm) were also
performed. The data analysis consists of debiassing, flat fielding and photometry using
SExtractor \citep{bertin} and was performed using the ATOM-specific pipeline processing \citep{Hauser04}.
The fluxes were determined using a 4$^{\rm ''}$ radius aperture and the flux
scale was fixed on 7 nearby stars from the USNO catalogue \citep{tosti}.

\begin{figure*}[!t]
   \centering \includegraphics[width=0.90 \textwidth]{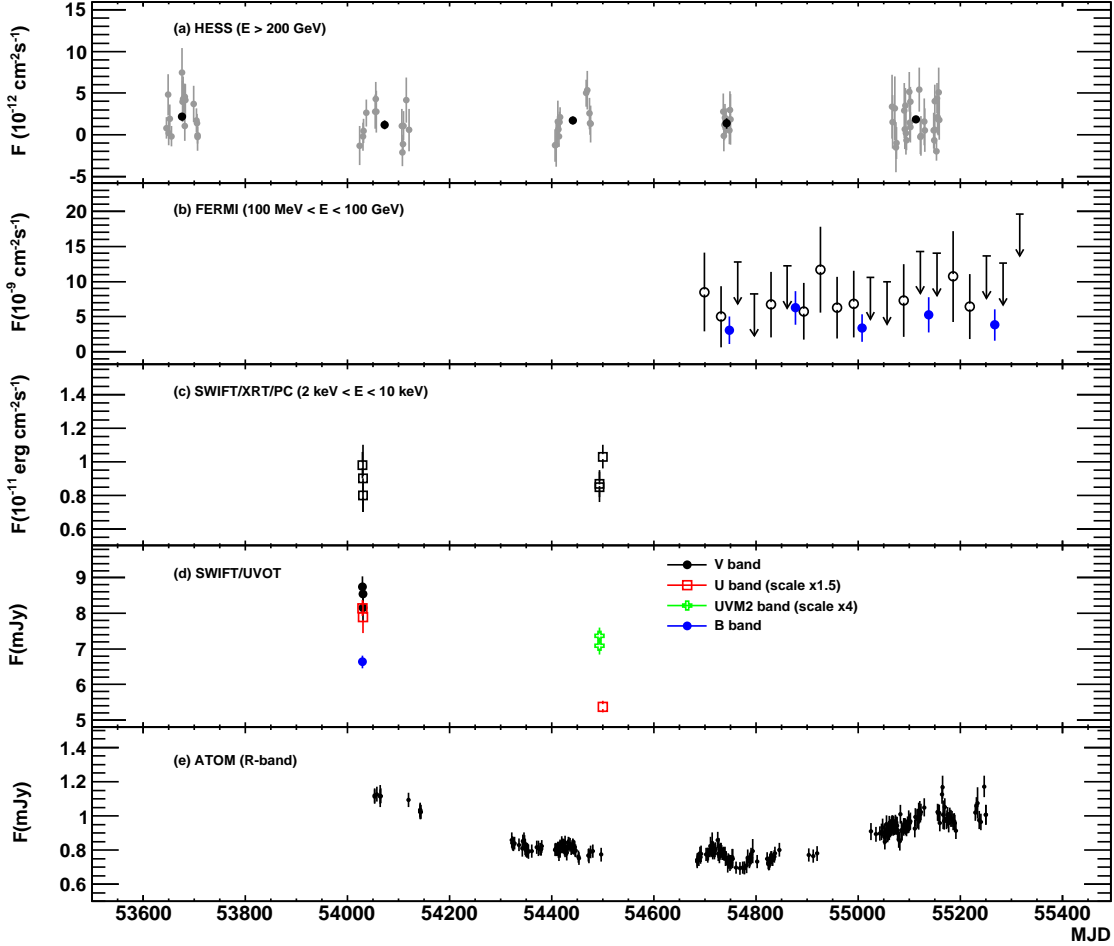}
   \caption{
Overall light curve of the multi-wavelength observations of \es covering the period from 2005 to 2010.
Night-wise (shaded) and year-wise (dark) light curves of 1ES~0414+009 by \hess (a).
The \textit{Swift}/XRT (c), UVOT (d), and ATOM (e) observations have a night-wise binning, while the \textit{Fermi} light curve (b) has 32.5 day binning with 95\% upper limits (black open circles) and 130 day binning (blue filled circles). Only statistical uncertainties of the flux points are shown.}
   \label{LC_MWL}
 \end{figure*}

The night-by-night light curve of the ATOM observations with the
R-filter is shown in the bottom panel of Fig.~\ref{LC_MWL}, where no
Galactic extinction correction was considered. There is no strong
variability in the light curve (the amplitude variations are less than
50\% overall) but a long-term trend can be seen, with a minimum in
2008 with respect to the other years. No change in (B-R) colour was
observed.

\section{Combined spectral fits} \label{spectral_fits}

\subsection{UVOT and XRT spectral fits} \label{uvot_xrt_fits}

To study the connection between the UV and X-ray components of the synchrotron emission, 
and to better estimate the position of its peak, 
a broad-band fit of the simultaneous UVOT and XRT data was performed in Xspec,
using the UVOT response matrices in the calibration database.
Only the B and UV filters were considered, to minimize the contamination from the host galaxy.
In Xspec, Galactic reddening and X-ray absorption are taken into account
using the models {\tt redden} \citep[updated with][]{odonnell96}
and {\tt wabs} for the two bands, respectively.

The UV-to-X-ray broadband spectra in all three epochs are well represented 
by a single log-parabolic component (Table~\ref{swift}), 
with values of the curvature which are typical for HBLs: around 0.1--0.2 \citep{tram07}.
The resulting estimate for the energy of the synchrotron peak is around 0.1 keV, overall,
with a slight shift (at $\sim3\sigma$ level) of the peak position from 0.07 to 0.17 keV 
between the 2006 and 2008 epochs.
The corresponding flux at the synchrotron peak is steady around ($1.25\pm0.10$)~$\times$~10$^{-11}$~\ergcms 
between 2006 and Jan 2008, decreasing slightly to ($1.01\pm0.11)$~$\times$10$^{-11}$~\ergcms on Feb 4, 2008.

\subsection{\textit{Fermi} and \hess spectral fits} \label{ebl}

By combining the \textit{Fermi} and \hess spectra it is possible to study the broadband properties of 
the $\gamma$-ray SED peak, once the effects of $\gamma-\gamma$ interactions
with photons of the diffuse EBL are taken into account.
At $z$ = 0.287, the intrinsic spectrum calculated from the VHE observations is strongly sensitive to the EBL density. 
Using the high EBL density given by direct estimates at 1-3~$\mu$m \citep[$\sim$20~nW$/$m${^2}$sr, see e.g.][]{nature}, the reconstructed spectrum is unrealistically hard ($\Gamma_{\rm int}\leq0.3$), but it softens to $\Gamma_{\rm int}\geq1.5$ for lower EBL densities.

Considered independently of other spectral information, therefore,
the \hess spectrum of \es further corroborates 
the upper limits on the EBL level previously derived from other HBLs \citep{nature,0347,1101}, 
indicating a low density around $\sim$1-3~$\mu$m  close to the lower bound
given by galaxy counts \citep{nature}.  However,
the measured spectrum of \es does not improve those upper limits, 
nor their statistical uncertainty.

For example, comparing with 1ES~1101-232 ($z = 0.186$) and
performing the same analysis described in \citet{nature}, where a scaled EBL spectral shape was
used, a higher EBL scaling factor P is required in this case as compared to
1ES~1101-232 (P0.58 vs P0.55) in order to give the same estimated intrinsic spectral index.
The higher redshift ($z = 0.287$ vs $z = 0.186$) provides a longer path-length over which the effect of EBL
absorption is integrated, thus giving a larger measurable effect of absorption of the
intrinsic spectrum (i.e. a smaller difference in EBL flux $\Delta$$F_{\rm EBL}$ is sufficient
to yield the same observed $\Delta\Gamma$). This longer leverage, however, is not enough
to compensate for the larger statistical error on the spectral slope ($\sim$0.35 vs $\sim$0.15). 

Therefore, the approach adopted here is to fix the EBL to the best current 
estimates \citep[e.g.][]{franceschini,primack},
and to investigate the resulting SED properties.
The EBL model assumed here \citep{franceschini} is close to the lower limits
from galaxy counts and compatible with the limits from VHE observations \citep{nature}.  
With this model, both the \textit{Fermi} and \hess absorption-corrected spectra are harder than 2,  
indicating a high-energy peak above few TeV. 
Both spectra have the same index, within statistical errors 
($\Gamma_{\rm HE}\simeq1.85\pm0.18$ {\em vs} $\Gamma_{\rm VHE, int}\simeq1.87\pm0.35$), 
indicating that the spectral properties do not change significantly between the HE and VHE bands. 
The typical uncertainty on the VHE spectral index due to the residual EBL uncertainty can be estimated to be 
in the range $[$+0.15,0.3$]$, scaling the Franceschini template between the level of galaxy counts 
and the current upper limits.
A single power-law model provides a good representation of the combined \textit{Fermi} and \hess spectra
from 1~GeV to 2~TeV, with index $1.86\pm0.06$ and $\chi^2/ndf$ = 4.7/7.

Because the \hess observations overlap only partially in time 
with the \textit{Fermi} data, variations in the flux could bias the combined spectral fitting 
of the average spectra towards harder or softer values.  
To quantify this, the \textit{Fermi} average spectrum during the \hess observing epochs 
($\sim$3.6 months) was compared to the total (20 months).

The differences are negligible
with respect to the statistical errors ($<$7\%), 
and the confidence contours in the flux-index plane of 
the total data set are fully enclosed in
 the respective contours of the \hess epoch. 
Thus, it is possible to conclude that no correction between the
\textit{Fermi} and \hess spectra is necessary.

\section{Discussion} \label{Discussion}

\begin{figure*}[!t]
  \centering \includegraphics[width=16cm]{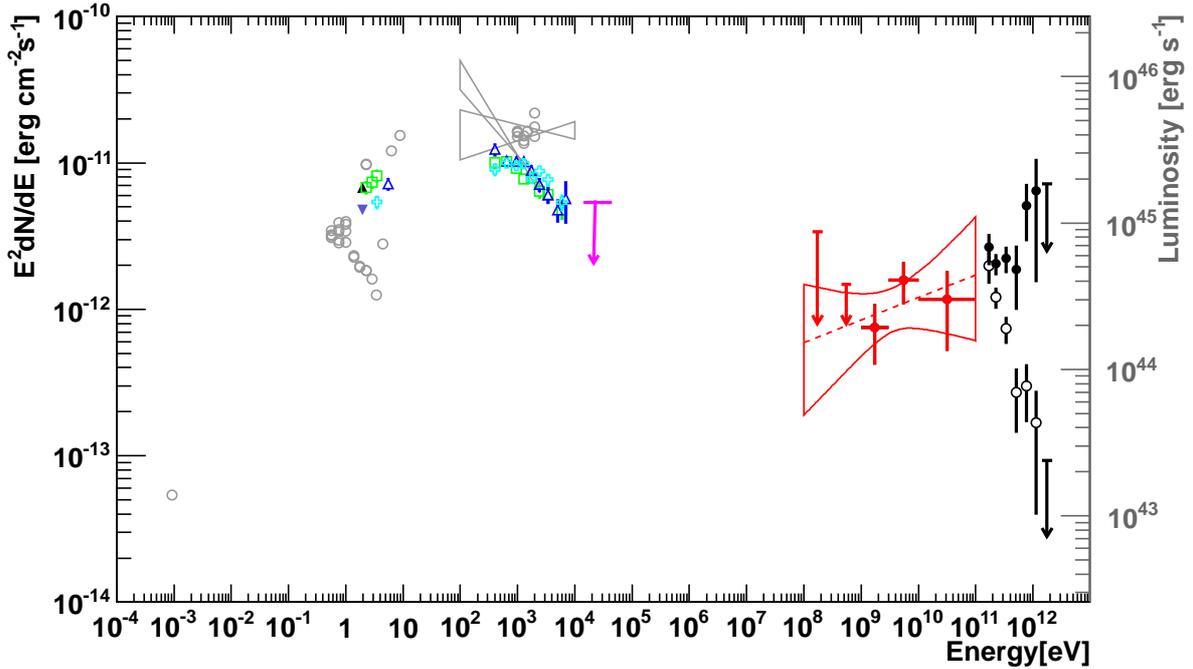}
   \caption{Average SED of \es from the observations reported in Fig.~\ref{LC_MWL} in the epoch 2005-2009.
H.E.S.S.: black filled/open circles with/without EBL correction, respectively. \textit{Fermi}: red full circles correspond to detections with $TS>9$, red downward-pointing arrows to 95\% upper limits. \textit{Swift}/BAT in 5 yrs:
magenta 99\% upper limit. \textit{Swift}/XRT\&UVOT: green open squares (Oct 2006), blue open triangles (Jan 2008), and cyan crosses (Feb 2008). ATOM: black triangle and blue inverted triangles corresponding to two different states, lower flux level in 2008 and higher in 2006, respectively, in the R and I bands. Gray points and butterflies are a collection of archival data from \citet{costa_ghisellini} and references therein.
}   \label{sed}
 \end{figure*}

The overall SED is shown in Fig.~\ref{sed}.
Unfortunately, the X-ray pointings are not strictly simultaneous with the \hess or \textit{Fermi} observing epochs.
(see Fig.~\ref{LC_MWL}).  
However, given the lack of strong variability
in all the observed bands, it can be reasonably assumed that the ATOM, \textit{Swift}, \hess and \textit{Fermi} data
provide together a good representation of the average SED properties of the source over
the time span of the observations,

These observations confirm \es as a typical HBL, 
with a synchrotron peak around 0.1~keV and where the X-ray band is dominated by the synchrotron 
emission of high-energy electrons. The higher-energy component, which is commonly interpreted as 
inverse Compton (IC) emission from the same population of electrons responsible for the 
synchrotron emission, was largely unexplored before the observations reported here.
The combined data indicate that a single, hard power law fits the intrinsic
spectrum over the combined HE-VHE $\gamma$-ray ranges, constraining the peak of the IC emission
to lie at energies greater than $\sim$1--2~TeV.  Thus, \es seems to belong to the subclass of HBLs 
characterized by Compton peak energies 
in the multi-TeV range, like 1ES\,1101-232 \citep{1101},
1ES\,0347-121 \citep{0347}, 1ES\,0229+200 \citep{0229,tav0229}, 
or 1ES\,1218+304 \citep{ver1218}.

Such high energies for the IC peak are  difficult to explain with
a standard one-zone SSC model,  
due to the decrease in the scattering efficiency in the Klein-Nishina regime 
and the decrease in the energy density of the seed photons available for scatterings in the Thomson regime.  
Both these effects tend to steepen the emitted $\gamma$-ray spectrum at VHE.  
In principle it would be possible to avoid this steepening by assuming that
the cooling of the electrons occurs only by IC scattering in the deep Klein-Nishina regime
\citep[for example, off a black-body-like spectrum of target photons, see][]{moderski2005}.
However, the resulting hardening of the electron distribution at high energies
(because high-energy electrons would cool slower than those at lower energies)
would necessarily imply a strong hardening of the synchrotron spectrum towards hard X-ray energies
\citep{moderski2005}. 
The \textit{Swift}/XRT and BAT data do not seem to corroborate this hypothesis,
though such an enhancement could in principle be located at even higher energies ($>100$~keV).
 
Therefore, to explain an SED such as that of \es (Fig.~\ref{sed}) in the context of a one-zone SSC model, 
one is drawn to consider a set of parameters for which the IC peak is obtained in the 
Thomson regime ($\gamma_{peak}h\nu'\leq m_e c^2/4$) 
\citep[see][for alternative scenarios]{intabs,boett}.
The branch of the electron distribution corresponding to the GeV-TeV band would then
correspond to the optical-UV band of the synchrotron spectrum 
(i.e. electron energies $\gamma<\gamma_{peak}$).

Adopting $h\nu_{\rm s}=0.1$ keV and $h\nu_{\rm c}=2$ TeV as the energies of the synchrotron and 
IC peaks respectively,
the Lorentz factor of the electrons at the peak would be  
$\gamma_{\rm p}=(3\nu_{\rm c}/4 \nu_{\rm s})^{1/2}\simeq1.2\times10^5$. 
The luminosity at these two energies can be estimated as 
$L_{\rm s}\sim10^{45.5}$ and $L_{\rm c}\sim10^{44.9}$~erg$/$s,
respectively, from the log-parabolic fit to the synchrotron peak and the power-law fit
to the $\gamma$-ray spectrum.
Performing an analysis of the independent SSC constraints in the (log ${\rm B}$ -- log $\delta$) plane
\citep[see e.g.][]{tavecchio98,guy}, with  these parameters, a 
solution in the Thomson regime can be found for $\delta\gtrsim50$ and $B\lesssim0.01$~G,
but only with a rather large emitting region (R~$\sim$~$10^{17}$~cm,
vs R$_S\sim$~$6\times10^{14}$~cm, the Schwartzschild radius of the putative Black Hole).
If this is the case, no variability faster than 1 day is expected, though 
this is below the sensitivity of current experiments at this
source's flux level.
Assuming instead a more typical size of the emitting region for HBLs, around R~$\sim1-2\times10^{16}$~cm,
rather extreme parameters would be required: $\delta>200$ and B~$<0.002$~G.

However, in both cases the synchrotron cooling is very slow, 
and the cooling time of the electrons at the synchrotron peak is much longer than the escape time
\citep[see e.g.][considering the electron escape velocity in the range 1-1/3$c$]{tavecchio98}.
The break in the electron spectrum corresponding to the SED peak, therefore,
cannot  be explained simply by radiative cooling (a ``cooling break")
or by an equilibrium between cooling and escape from the source, as typically assumed 
for BL~Lac objects with much lower Compton-peak energies, like PKS~2155-304, 
Mkn~421 or Mkn~501 \citep[e.g.][]{tavecchio98,guy,kraw2002,gcc}.
Furthermore,  
the slow cooling requires the kinetic energy of the jet to be increased significantly in order to
compensate for the low efficiency of the radiative emission.
It should be noted that the \textit{Fermi} and \hess spectra provide only a lower limit
on the energy and luminosity of the IC peak; if the true source values are actually higher, 
even more extreme parameters would be required.
The problems just described are 
a general issue for the class of TeV BL~Lac objects characterized by a hard TeV spectrum,
an IC peak at very high energies and a large separation between synchrotron and IC peak frequencies.
It might indicate either a fundamental difference in the 
physical jet conditions with respect to the other HBLs, or the need for an emission mechanism
different from that of homogeneous SSC models. 

As a final caveat, however, it should be noted that the statistical uncertainty on the $\gamma$-ray spectra 
is rather large, especially at VHE ($\Gamma_{\rm VHE, int} = 1.87\pm0.35$). 
If the true VHE spectrum were $\geq1 \sigma$ steeper
than the H.E.S.S. measurement (i.e. $\Gamma_{\rm VHE, int}=2.2$), 
the IC component would have a peak energy around 100--200 GeV,
and a slightly lower luminosity. This would make the \es SED more similar to a typical HBL
and would be easier to explain with standard SSC parameters.

\section{Conclusion} \label{conclusion}

The \hess array of Cherenkov telescopes has been used to detect significant VHE $\gamma$-ray emission in
the direction of \esnospace, a distant BL Lac objects of known redshift.  
The overall flux above 200 GeV from 2005 to 2009 shows no variability on any time-scale and 
is on average very low (around 0.6\% Crab Nebula flux), making \es one of the faintest 
extra-galactic sources detected in the TeV domain.
This source was detected with the \textit{Fermi} LAT in the first 20 months of its operation between
2008 and 2010 and was very faint in the HE domain too.

The VHE spectrum measured using \hess is consistent with the current limit of the EBL and confirms the
low level of EBL in the few $\mu$m range as derived from galaxy counts. 
The HE and VHE spectra (absorption corrected with an EBL model close to the lower limits) 
show a best-fit power law with an index harder than 2, indicating that \es can  
be classified as a hard-TeV BL~Lac object. The overall SED of this source is averaged 
over five years with no strict simultaneity between the \textit{Swift} and \hess observations. 
With this caveat, the properties of the SED 
--in particular an IC peak energy above 1--2~TeV--
are difficult to explain in the framework of a pure one-zone SSC model, 
unless using unusual values for the main parameters.

Future observations of \es are strongly motivated, in particular with the next generation 
of Cherenkov telescope arrays such as the Cherenkov Telescope Array (CTA). 
With a tenfold increase in sensitivity with respect to the 
current Cherenkov experiments, 
CTA will allow better measurements of the VHE spectrum of faint sources, 
especially at high energies, provide better constraints of the IC peak, and will help to
address the physics issues posed by the hard-TeV BL~Lac objects.

\section*{Acknowledgements}

{\small The support of the Namibian authorities and of the University of Namibia
in facilitating the construction and operation of \hess is gratefully
acknowledged, as is the support by the German Ministry for Education and
Research (BMBF), the Max Planck Society, the French Ministry for Research,
the CNRS-IN2P3 and the Astroparticle Interdisciplinary Programme of the
CNRS, the U.K. Science and Technology Facilities Council (STFC),
the IPNP of the Charles University, the Polish Ministry of Science and 
Higher Education, the South African Department of
Science and Technology and National Research Foundation, and by the
University of Namibia. We appreciate the excellent work of the technical
support staff in Berlin, Durham, Hamburg, Heidelberg, Palaiseau, Paris,
Saclay, and in Namibia in the construction and operation of the
equipment.\\
The \textit{Fermi}-LAT Collaboration acknowledges generous ongoing support
from a number of agencies and institutes that have supported both the
development and the operation of the LAT as well as scientific data analysis.
These include the National Aeronautics and Space Administration and the
Department of Energy in the United States, the Commissariat \`a
l'Energie Atomique
and the Centre National de la Recherche Scientifique / Institut
National de Physique
Nucl\'eaire et de Physique des Particules in France, the Agenzia
Spaziale Italiana
and the Istituto Nazionale di Fisica Nucleare in Italy, the Ministry
of Education,
Culture, Sports, Science and Technology (MEXT), High Energy Accelerator Research
Organization (KEK) and Japan Aerospace Exploration Agency (JAXA) in Japan, and
the K.~A.~Wallenberg Foundation, the Swedish Research Council and the
Swedish National Space Board in Sweden. Additional support for science
analysis during the operations phase is gratefully acknowledged from
the Istituto Nazionale di Astrofisica in Italy and the Centre National
d'\'Etudes Spatiales in France.\\
This research has made use of data obtained from 
the High Energy Astrophysics Science Archive Research Center 
(HEASARC), provided by NASA's Goddard Space Flight Center.
}

\end{document}